\documentclass{webofc}
\usepackage[varg]{txfonts}   
\usepackage{hyperref}
\usepackage{url}
\hypersetup{colorlinks=true,citecolor=blue,urlcolor=blue,linkcolor=blue}
\begin{document}

\title{Pre-equilibrium charm quark dynamics and their impact on D-Meson observables}

\author{
    \firstname{Manu} \lastname{Kurian}\inst{1}  \and 
    \firstname{Mayank} \lastname{Singh}\inst{2} \and
    \firstname{Charles} \lastname{Gale}\inst{3} \and
    \firstname{Sangyong} \lastname{Jeon}\inst{3} \and    
    \firstname{Bj\"orn} \lastname{Schenke}\inst{4}\fnsep\thanks{\email{bschenke@bnl.gov}}        
}

\institute{Department of Physics, Indian Institute of Technology (Indian School of Mines) Dhanbad, Jharkhand 826004, India 
\and
        Department of Physics and Astronomy, Vanderbilt University, Nashville, TN 37240, USA
\and
        Department of Physics, McGill University, 3600 University Street, Montreal, QC H3A 2T8, Canada
\and
        Physics Department, Brookhaven National Laboratory, Upton, NY 11973, USA
          }

\abstract{We study the impact of pre-equilibrium evolution on the charm quark $R_{AA}$ and $v_2$ in Pb+Pb collisions at $\sqrt{s_{NN}}$ = 5.02 TeV. We observe that there is significant diffusion in the pre-equilibrium evolution, but there is no measurable effect on the final state observables.
}
\maketitle
\section{Introduction}

\label{intro}
Heavy quarks (charm and bottom) are produced at the earliest stages of heavy-ion collisions owing to their large masses. They take longer to thermalize than their lighter counterparts, and their motion through the quark-gluon plasma can be described as a Brownian motion within the Langevin or Fokker-Planck frameworks \cite{mustafa1,moore1}.

It has been proposed that heavy-quark observables can help provide insight into the stage of pre-equilibrium bulk evolution \cite{das1,carrington1}. While the lifetime of the pre-equilibrium stage is relatively short, it comprises significantly higher energy densities. As a result, heavy quarks are postulated to have much stronger interactions in the earliest stages, which can leave an imprint on final-stage heavy flavor observables.

Here, we summarize the consequences of pre-equilibrium interactions on two charm observables: the nuclear modification factor ($R_{AA}$) and the elliptical anisotropy coefficient ($v_2$) of D-mesons. This is based on our study where we describe the light and heavy flavor observables using the IP-Glasma + MUSIC + UrQMD + MARTINI framework \cite{Singh:2025duj}.

\section{Model Setup}

\subsection{Bulk medium}

The initial stage of the bulk medium is modeled using IP-Glasma \cite{schenke1}. Nucleon positions in the colliding nuclei are sampled from the Woods-Saxon distribution, and the binary collision positions are identified using the nucleon-nucleon cross-sections. Color charges are sampled provided the nucleon position information, and gluon fields are determined and evolved using the classical Yang-Mills equations. We track the energy density at different space-time points during the evolution.

Starting from $\tau = 0.4$ fm, the medium is evolved using the relativistic viscous hydrodynamic code MUSIC \cite{schenke2}, and a constant energy density freezeout at the 0.18 GeV/fm$^3$ energy density hypersurface is saved. It is used to compute hadrons, which are sampled using the iSS package \cite{shen1}. These hadrons then undergo scatterings and decays in UrQMD \cite{bass1}. The model parameters are the same as in \cite{schenke3}.

\subsection{Charm production}

Charm quarks are produced at the binary collision positions determined earlier. We use PYTHIA 8.2 \cite{sjostrand1} to sample $p-p$, $p-n$, and $n-n$ collisions to generate heavy quark $Q\bar{Q}$ pairs from initial hard scatterings and gluon splittings. The medium modification to charm production rates is not considered. Nuclear modification to the parton distribution functions are accounted for by using the distributions from EPS09 \cite{eskola1}. We allow a finite formation time of $1/m_c$ in the local rest frame of the charm quark where $m_c$ is the charm mass.

\begin{figure}
\centering
\includegraphics[width=0.45\textwidth]{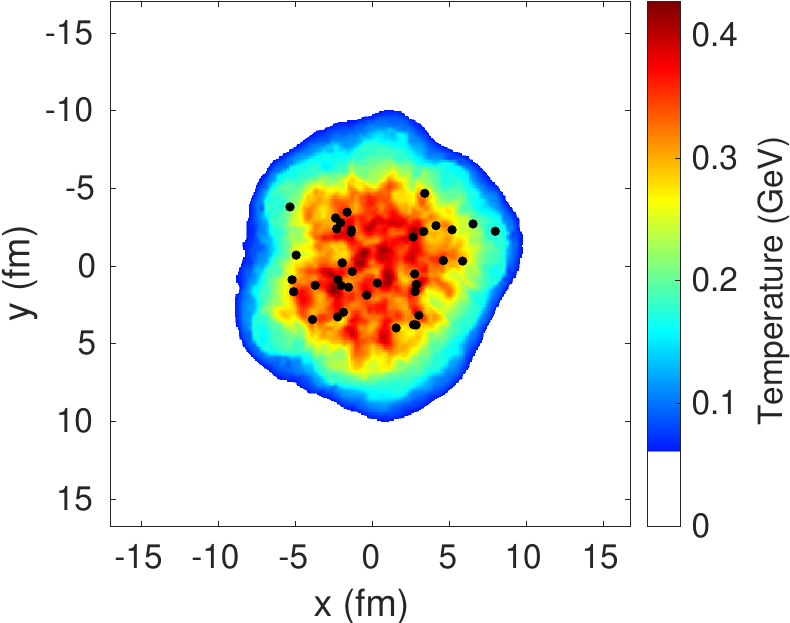}
\caption{Charm quark positions in the background medium at $\tau =$ 1.5 fm at midrapidity.}
\label{fig-1} 
\vspace{-0.5cm}

\end{figure}

\subsection{Charm propagation}

Charm quark propagation through the pre-equilibrium and the hydrodynamic medium is treated in the Langevin framework within MARTINI \cite{schenke4,young1}. The discretized version of the Langevin equation in the local rest frame of the medium is given by
\begin{align}
   &dp_i=-\eta(|{\bf p}|) p_i\, dt+ \xi_i ({\bf p})\, {dt},\nonumber\\
   &\langle \xi_i (t)\xi_j (0)\rangle = \Big(\delta_{ij}-\frac{p_ip_j}{|{\bf p}|^2}\Big)\,\kappa_T(|{\bf p}|)+\frac{p_ip_j}{|{\bf p}|^2}\,\kappa_L (|{\bf p}|),
\end{align}
where $dp_i$ is the change in the $i$-th component of momentum $p_i$ of the charm quark. The drag coefficient is given by $\eta$ while the transverse and longitudinal diffusion coefficients are denoted as $\kappa_T$ and $\kappa_L$. Here, $\xi$ is the stochastic term whose strength is determined by the size of the diffusion coefficients. In this work, we chose $\kappa_T = \kappa_L = \kappa$.

The spatial diffusion coefficient $D_s$ is obtained from 2+1 flavor lattice QCD \cite{hotqcd1}. It is related to $\eta$ and $\kappa$ at zero relative momentum using Einstein relations. We parametrize the momentum dependence of $\eta$ and $\kappa$ from perturbative calculations \cite{singh1} and use the lattice results to fix the values at zero momentum. 

Obtaining drag and diffusion coefficients in the pre-equilibrium phase requires defining a temperature in this regime. We assume isotropy and use the ideal gas equation of state with gluon degrees of freedom to define the temperature here. This is a simplification aimed at getting a rough estimate of the size of the phenomenological effect of pre-equilibrium charm interactions. The spatial diffusion coefficient can be obtained from zero-flavor lattice calculations \cite{banerjee1}. When adjusting for different critical temperatures, the temperature dependence of $D_s$ for zero-flavor and 2+1 flavor lattice calculations is consistent \cite{HotQCD}.
\begin{figure*}
\centering
\includegraphics[width=0.45\textwidth]{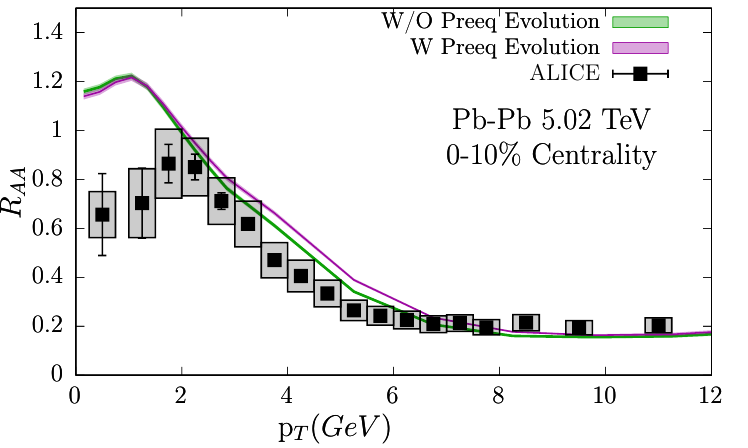}
\includegraphics[width=0.45\textwidth]{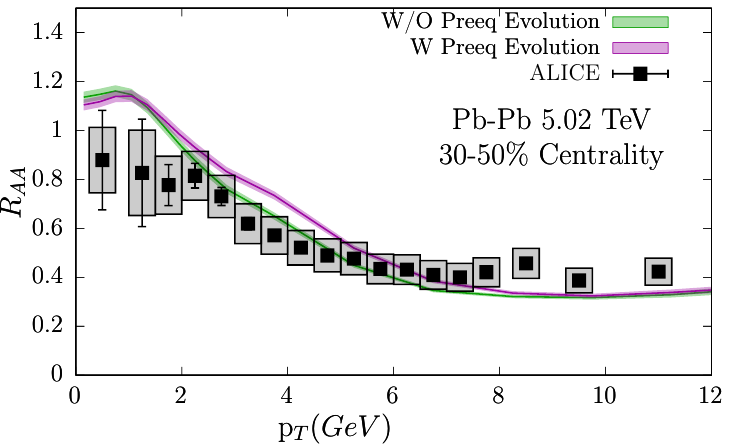}
\caption{Impact of charm quark pre-equilibrium evolution on D-meson $R_{AA}$. Experimental data taken from~\cite{ALICE:2021rxa,ALICE:2020iug}.}
\label{fig-2}  
\vspace{-0.5cm}

\end{figure*}
\begin{figure*}
\centering
\includegraphics[width=0.45\textwidth]{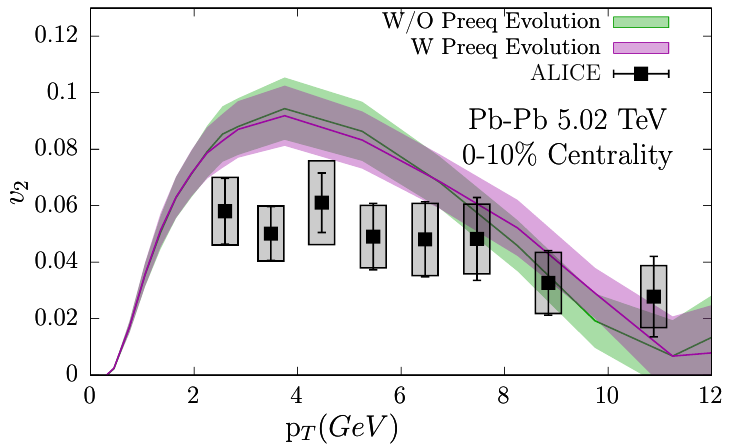}
\includegraphics[width=0.45\textwidth]{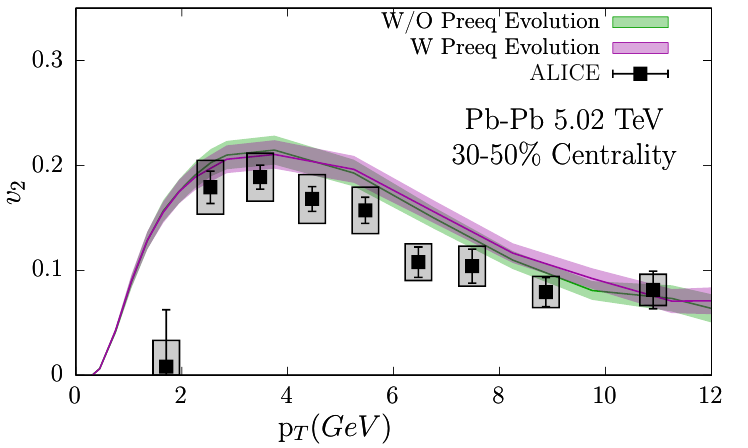}
\caption{Impact of charm quark pre-equilibrium evolution on D-meson $v_2$. Experimental data taken from~\cite{ALICE:2021rxa,ALICE:2020iug}.}
\label{fig-3}   
\vspace{-0.5cm}

\end{figure*}

\subsection{Hardronization}

Charm quarks are hadronized when their surrounding medium temperature falls to 165 MeV. We hadronize using a hybrid coalescence + fragmentation approach. Coalescence is done using the Duke coalescence model \cite{cao1}. We include the D-mesons and charmed baryons $\Lambda_c$, $\Sigma_c$ $\Omega_c$ and $\Xi_c$. The overall coalescence probability is normalized by a free parameter. For this work, the parameter is chosen such that the net coalescence probability for a quark at rest in the medium is 0.2. Charm quarks, which do not coalesce, are allowed to fragment where the momentum fraction is generated using the Peterson fragmentation function \cite{Peterson:1982ak}.

\section{Results and discussion}

Figure \ref{fig-1} shows the charm quark positions at mid-rapidity at $\tau =$ 1.5 fm. We calculate the D-meson observables $R_{AA}$ and $v_2$ for two centralities 0-10\% and 30-50\%. Figures \ref{fig-2} and \ref{fig-3} depict $R_{AA}$ and $v_2$ of D-meson for these two centralities with and without the presence of pre-equilibrium evolution. We observe that the pre-equilibrium evolution of the charm quark slightly modifies the $R_{AA}$, especially in the $p_T$ range $2-6$ GeV, while its effect on $v_2(p_T)$ remains negligible. Furthermore, we analyzed the dependence of the formation time of charm quarks and the uncertainty band in the lattice estimation of $2\pi D T$ on $R_{AA}$.  We observe that the formation time of the charm quark in the medium plays a crucial role in the high $p_T$ regime of the D-meson $R_{AA}$. Additionally, uncertainties in the diffusion coefficient estimation introduce noticeable effects on the $R_{AA}$, particularly in the high transverse momentum region.\\

\textbf{Acknowledgments} This research used the resources of the National Energy Research Scientific Computing Center, a DOE Office of Science User Facility
supported by the Office of Science of the U.S. Department of Energy
under Contract No. DE-AC02-05CH11231 using NERSC awards NP-ERCAP0033578 and NP-ERCAP0032155. M.K. acknowledges the Special Postdoctoral Researchers Program of RIKEN, the Faculty Research Scheme (FRS project number: MISC 0240) at IIT (ISM) Dhanbad, and the Department of Science and Technology (DST), Govt. of India, for the INSPIRE-Faculty award (DST/INSPIRE/04/2024/001794). This work is supported by the U.S. Department of Energy, Office of Science, Office of Nuclear Physics, under DOE Contract No.~DE-SC0024711 (M.S.) and~DE-SC0012704 (B.P.S.) and within the framework of the Saturated Glue (SURGE) Topical Theory Collaboration (B.P.S.). M.S. is also supported by the Vanderbilt University. S.J. and C.G. are supported by the Natural Sciences and Engineering Research Council of Canada under grant numbers SAPIN-2024-00026 and SAPIN-2020-00048, respectively.

\end{document}